\documentclass[12pt]{iopart}

\usepackage{graphicx}% Include figure files
\DeclareGraphicsExtensions{.eps,.ps,.eps.gz,.ps.gz,.eps.Z}
\begin{document}

\title [Spin droplets in realistic quantum Hall devices]{Stability of spin droplets in realistic quantum Hall devices}

\author{H. Atci$^1$$^,$$^2$, U. Erkarslan$^3$, A. Siddiki$^1$ and E. R{\"a}s{\"a}nen$^4$$^,$$^2$}
\address{$^1$ Physics Department, Faculty of Science, Istanbul University, 34134 Vezneciler Istanbul, Turkey}
\address{$^2$ Nanoscience Center, Department of Physics, University of Jyv{\"a}skyl{\"a}, FI-40014 Jyv{\"a}skyl{\"a}, Finland}
\address{$^3$ Department of Physics, Faculty of Science, Mugla University, 48170 K\"otekli-Mu\~gla, Turkey}
\address{$^4$ Department of Physics, Tampere University of Technology, FI-33101 Tampere, Finland}
\ead{huseyinatci@gmail.com}

\begin{abstract}
We study the formation and characteristics of ``spin droplets'',i.e., compact spin-polarized configurations in the highest occupied Landau level, in an etched quantum Hall device at filling factors $2\leq\nu\leq3$. The confining potential for electrons is obtained with self-consistent electrostatic calculations on a GaAs/AlGaAs heterostructure with experimental system parameters. Real-space spin-density-functional calculations for electrons confined in the obtained potential show the appearance of stable spin droplets at $\nu\sim 5/2$. The qualitative features of the spin droplet are similar to those in idealized (parabolic) quantum-dot systems. The universal stability of the state against geometric deformations underline the applicability of spin droplets in, e.g., spin-transport through quantum point contacts.
\end{abstract}

%Uncomment for PACS numbers title message
% Keywords required only for MST, PB, PMB, PM, JOA, JOB? 
%\vspace{2pc}
%\noindent{\it Keywords}: Article preparation, IOP journals
% Uncomment for Submitted to journal title message
%\submitto{\JPA}
% Comment out if separate title page not required
\maketitle

\section{Introduction}
Recent research on semiconductor quantum dots (QDs) 
has strongly focused on spin effects due to 
experimental breakthroughs in the initialization,
control, and readout of spin states that have
decoherence times up to milliseconds \cite{label1}.
Consequently, QDs are among the leading candidates 
for solid-state qubit design. On the other hand,
studies on few-electron QDs in strong magnetic fields
have shown interesting similarities to partially 
or fully spin-polarized quantum Hall (qH) states in 
the two-dimensional (2D) electron gas \cite{label2} (2DEG). For example, the filling factor $\nu=1$ in the 2DEG corresponds to a ``maximum-density droplet'' \cite{label3} in a few-electron QD. As another example, the Laughlin wave function \cite{label4} -- describing the filling factor $\nu=1/3$ state in the 2DEG -- was found to have $98\%$ overlap with the corresponding few-electron QD (three vortices per 
electron) \cite{label5}.

The analogy between the 2DEG and QDs applies further
to the $\nu=5/2$ state, although a direct comparison
is more complex. A. Harju {\em etal.} \cite{label6} showed that half-integer filling-factor states in QDs 
correspond to a situation where the highest occupied
Landau level (LL) is fully spin-polarized, whereas the
lower LLs are spin-compensated. Later on, the spin 
polarization of the highest LL is shown with three independent 
spin-blockade experiments \cite{label7, label8}.
Interestingly, these ``spin droplets'' were found to
form only when the number of confined electrons 
exceeded $\sim 30$ (Ref. \cite{label7}). 
In this respect, spin droplets are collective states induced
by a high density of states close to the Fermi level
that might lead to collective spin polarization.
On the other hand, the correspondence between the highest-LL 
spin droplets and the candidates of the $\nu=5/2$ state in the 
2DEG, such as the Pfaffian wave function, was found to be 
ambiguous \cite{label9}.

Although numerical studies have shown the appearance of spin droplets also in quantum rings \cite{label10}, there is no systematic investigation on the {\em stability} of those states. This issue is of fundamental importance when the spin droplets will be applied in more general qH devices such as quantum point contacts \cite{label11, label12}. In this work we will focus on this aspect of stability and perform a thorough theoretical investigation on the formation of spin droplets in a realistic qH device starting from actual device geometry and parameters. Our calculations show that the spin-droplet formation in a realistic device is very similar to that in ``idealized'', i.e., parabolically (harmonically) confined QDs. Thus, our finding confirms the high stability of spin droplets against geometric deformations. This might have significant consequences for the applicability of those states in, e.g., spintronics. 

The paper is organized as follows. In Sec.~\ref{methods}
we first perform self-consistent electrostatic calculations to extract the confining potential for electrons in a realistic qH device.
Then we present the many-electron Hamiltonian and our
spin-density-functional-theory \cite{label13} (SDFT) approach to
calculations in the qH regime. Our results for total energies,
total and spin densities, and chemical potentials are presented
in Sec. \ref{results}. The paper is summarized in Sec. \ref{summary}.

%============================================================
%============================================================
%============================================================

\section{Methods}\label{methods}

\subsection{Quantum-dot structure and the confining potential}\label{conf}

\begin{figure}[htb]
\centering
\includegraphics[width=0.65\columnwidth]{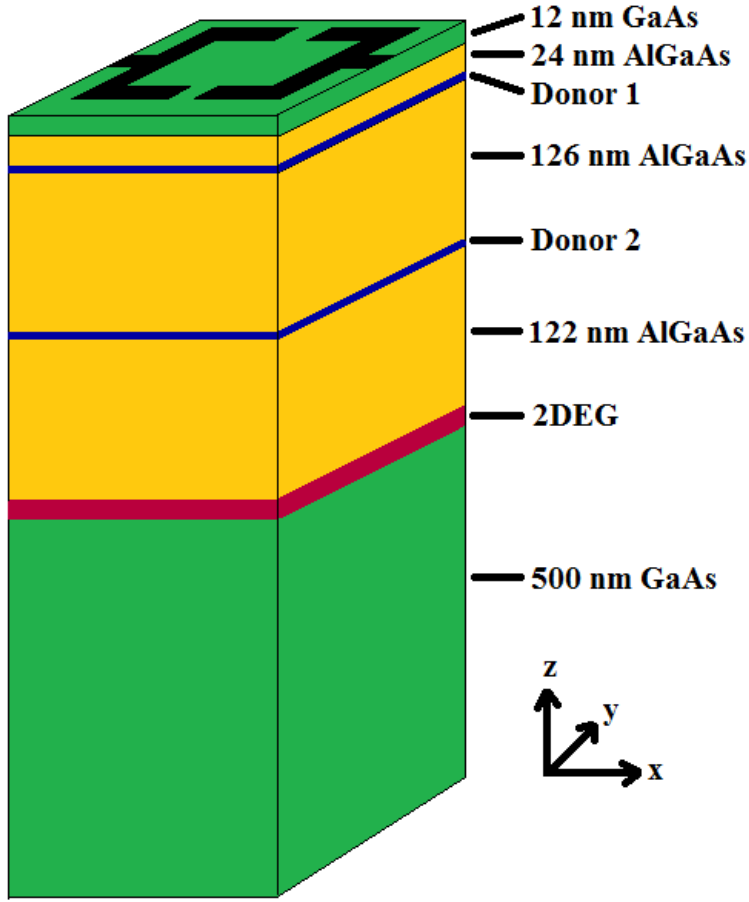}
\caption{Structure of the qH
device applied in the calculation of the confinement potential.
The material parameters are from Goldman \cite{label14}.}
\label{figure1}
\end{figure}

We consider a two layer $\delta$-Si doped GaAs/AlGaAs heterostructure provided by Goldman \cite{label14} and visualized in Figure \ref{figure1}. Here, the crystal is grown on a GaAs
substrate and the 2DEG is formed at the interface of the GaAs/AlGaAs
heterostructure located 284 nm below the surface.
The donor layers located 122 nm and 248 nm above the 2DEG
have surface densities $2.5\times10^{15}$ m$^{-2}$ 
and $1.7\times10^{16}$ m$^{-2}$, respectively.
The geometry of the QD is shown on top of the GaAs/AlGaAs 
heterostructure in Figure \ref{figure1}. The actual sample is obtained 
with etching 80 nm from the surface (in the $z$ direction). 
The 2D dimensions on the $xy$ plane are $L_{x}=L_{y}=2550$ nm. 

The QD sample is mapped on a matrix with $128\times128\times32$ 
mesh points. The confinement profile for the electrons trapped
in the 2DEG, i.e., for our QD, is calculated by solving the Poisson 
equation in 3D self-consistently within the Thomas-Fermi 
approximation \cite{label15, label16}. In the numerical calculation we apply a fourth-order nearest neighbor approximation and a 3D Fourier transformation. Here we apply open boundary conditions, since the heterostructure is embedded in a dielectric surrounding with a very small dielectric constant compared to the heterostructure.

The obtained confining potential is shown in Figure \ref{figure2}. It is noteworthy that the potential is not rotationally symmetric in contrast with the commonly used parabolic approximation for both lateral and vertical semiconductor QDs \cite{label17}. Moreover, the shallow etching leads to a relatively smooth potential, although the 
slope of the confinement is still considerably steeper than in a parabolic QD.

In the following we may consider the obtained confining potential in Figure \ref{figure2} as a trap for transported electrons forming a QD inside the device.
Alternatively, we can think that -- by adjusting 
the gate and bias voltages -- the 2DEG inside the QD is emptied 
from conduction-band electrons one-by-one. 
It is important to note that in the following
calculations we focus on such a QD-like system with a few dozen electrons instead of the bulk 2DEG in the qH regime. Our particular task is to investigate the effect of a relatively sharp and rotationally non-symmetric potential on the formation of spin-droplets in QDs at $\nu\sim 5/2$.

\begin{figure}[htb]
\centering
\includegraphics[width=1\columnwidth]{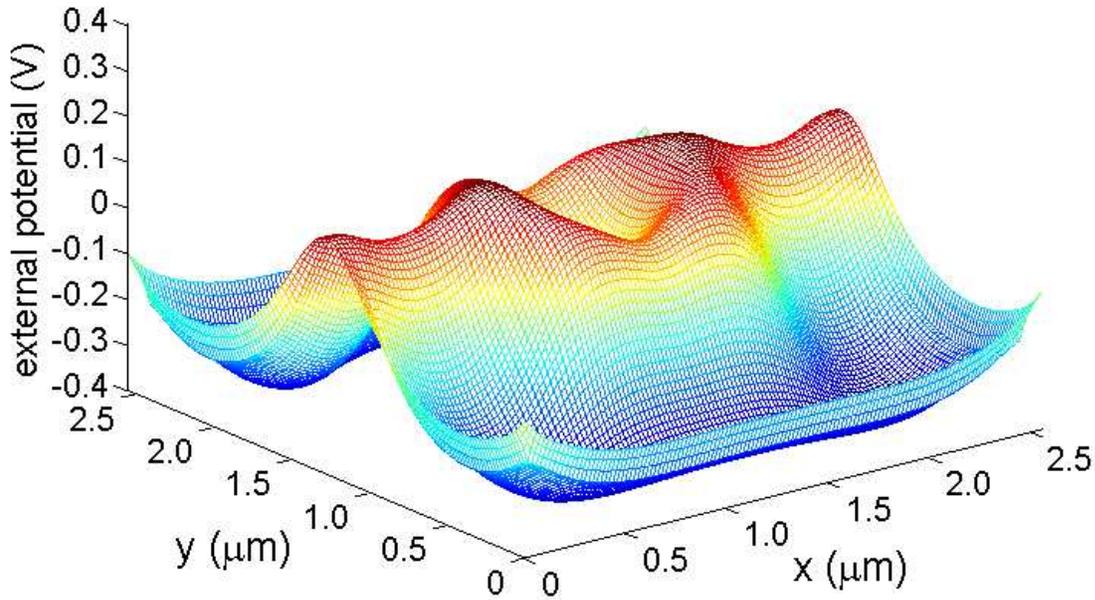}
\caption{Confining potential for electrons in the QD 
obtained with electrostatic calculations for the GaAs/AlGaAs 
heterostructure shown in Figure \ref{figure1}.}
\label{fig2}
\end{figure}

%============================================================
%============================================================

\subsection{Hamiltonian and many-electron calculations}\label{sdft}

Electrons in the QD are described by the Hamiltonian
\begin{equation}
H=\sum^N_{i=1}\left[\frac{({\bf p}_i+e {\bf A} )^2}{2 m^*}
+V_c({\mathbf r}_i)+g^{*}\mu_{B}S_{z,i}\right] + 
\frac{e^2}{4\pi \epsilon} \sum_{i<j}
\frac{1}{{\mathbf r}_{ij}},
\label{hamiltonian}
\end{equation}
where $N$ is the number of electrons, $V_{c}$ is the confining 
potential obtained in the previous section (Figure \ref{figure2}), and
${\bf A}$ is the vector potential of the homogeneous magnetic 
field oriented perpendicular to the QD plane. We consider the
conventional effective-mass approximation \cite{label17} with GaAs material
parameters $m^{*}=0.067m_{e}$, $\epsilon=12.7\epsilon_{0}$, and
$g^{*}=-0.44$ for the gyromagnetic ratio. 

As we consider $N\approx 48$ confined electrons, the many-electron
Hamiltonian is not solvable numerically exactly. The non-circular
geometry sets additional limitations. Hence, we use SDFT \cite{label13} 
that has been shown to produce reliable results when
compared with quantum Monte Carlo (QMC) calculations \cite{label7, label9}.
An explicit comparison between SDFT and QMC results for
spin-state energies can be found in Fig. 8 of Ref. \cite{label9}.
For the exchange and correlation we use the 2D
local-spin-density approximation (LSDA) with the parametrization
of Attaccalite {\em et al.} \cite{label18} for the
correlation. In view of the previous 
works, \cite{label7, label8, label9} we rely 
on the computationally efficient LSDA
despite recent progress in the development of alternative
and more accurate density functionals
for 2D systems \cite{label19, label20, label21}.

Our real-space SDFT approach with the {\tt octopus} 
code \cite{label22} allows the use of the confining potential
in Figure \ref{figure2} as a direct input. We calculate the 
total energies and spin densities for different spin 
configurations, respectively, and determine the ground state 
as the solution with the lowest energy. It should be noted
that in lack of rotational symmetry the angular momentum
is not a good quantum number. However, the (approximate) 
angular momenta of the effective single-electron (Kohn-Sham) 
states enable us to determine the occupations of different
LLs (see below).

%============================================================
%============================================================

\section{Results}\label{results}

\subsection{Total energies}

In the results presented below the magnetic-field range has been
selected such that the regime at filling factors $2\leq\nu\leq3$
is covered. The filling factor in a QD can be approximated 
by $\nu\approx 2N/N_{\rm 0LL}$, where $N_{\rm 0LL}$ is the number
of electrons in the lowest LL \cite{label7}. 
We point out that this approximation is valid only at $\nu\geq 2$,
whereas otherwise a good estimate is given by $\nu=N/N_{\Phi_0}$, 
where $N_{\Phi_0}$ is the number of flux quanta $\Phi_0=e/h$.
Detailed comparison between different definitions for the
filling factor in confined systems is given in Ref. \cite{label10}.

\begin{figure}
\includegraphics[width=1\columnwidth]{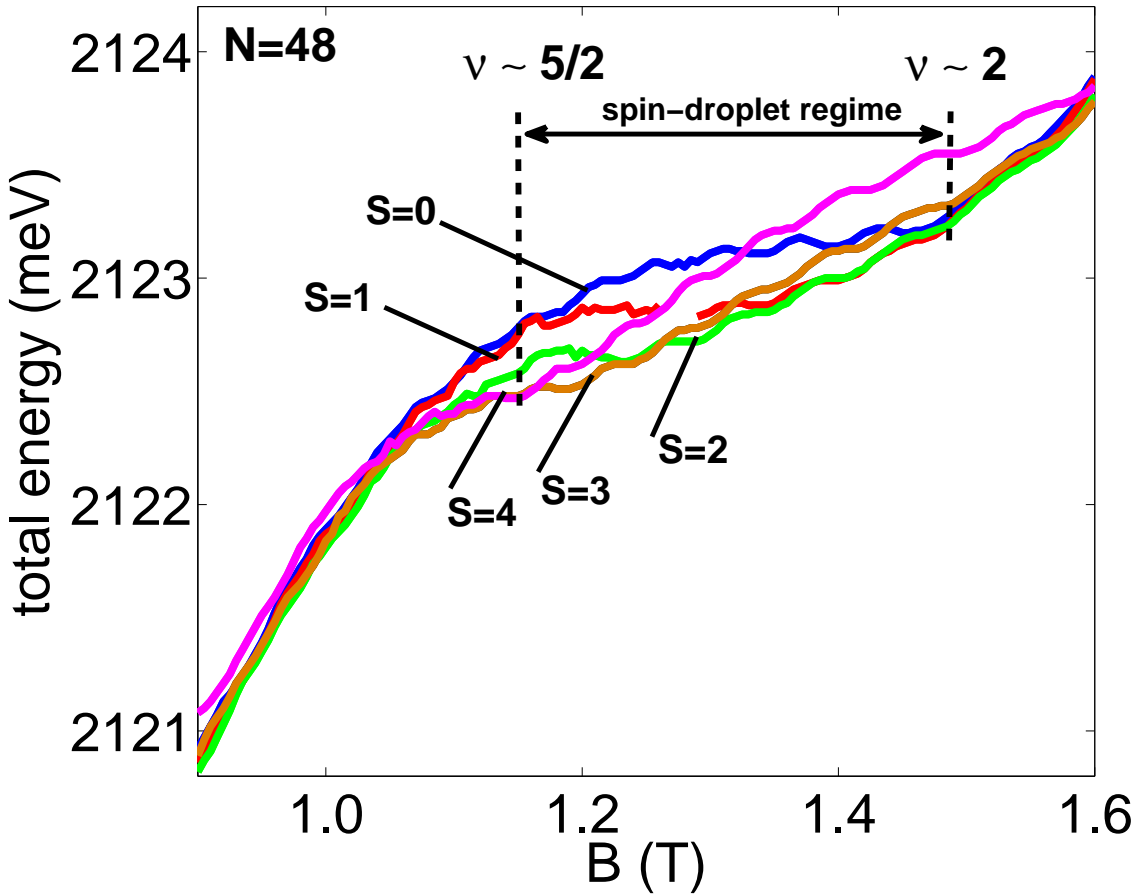}
\caption{Total energies of different spin states as a function of the magnetic field in a 48-electron QD calculated with SDFT 
in the spin-droplet regime ($2\leq\nu\leq 3$).}
\label{figure3}
\end{figure}

The total energy can be written
as a sum $E_{\rm tot}=E_{\rm kin}+E_{\rm ext}+E_H+E_{xc}$, where
$E_{\rm kin}$ is the (Kohn-Sham) kinetic energy, 
$E_{\rm ext}=\int\,d{\mathbf r}\,\rho({\mathbf r})V_c({\mathbf r})$
is the external confinement energy, $E_H$ is the classical
electrostatic (Hartree) interaction energy, and $E_{xc}$ is
the exchange-correlation energy accounting for quantum-mechanical 
interaction energy components beyond $E_H$.
In Figure \ref{figure3} we show the total energies of different
spin states $S$ as a function of the magnetic field for a 48-electron 
QD. The points of $\nu \sim 5/2$ and
$\nu \sim 2$ according to the above definition are marked
by dashed lines. At $B\leq 1.1$ T the states with different $S$
are almost degenerate until a distinctive energy gap opens up between
the higher and lower $S$. The maximum ground-state spin is $S=4$, 
and as $B$ is increased the polarization of the QD
gradually decreases: at $B \geq 1.5$ T the ground state has
$S=1$, and the unpolarized $S=0$ state is again very close in energy.

Overall, the behavior of $E_{\rm tot}$ in Figure \ref{figure3} 
is very similar to that of parabolic QDs \cite{label7}.
In the following we analyze the corresponding spin densities
and chemical potentials in detail in order to characterize 
the expected spin-droplet formation.

%============================================================
%============================================================

\subsection{Spin densities at $\nu\sim 2$ and $\nu\sim 5/2$}

\begin{figure}
\includegraphics[width=1\columnwidth]{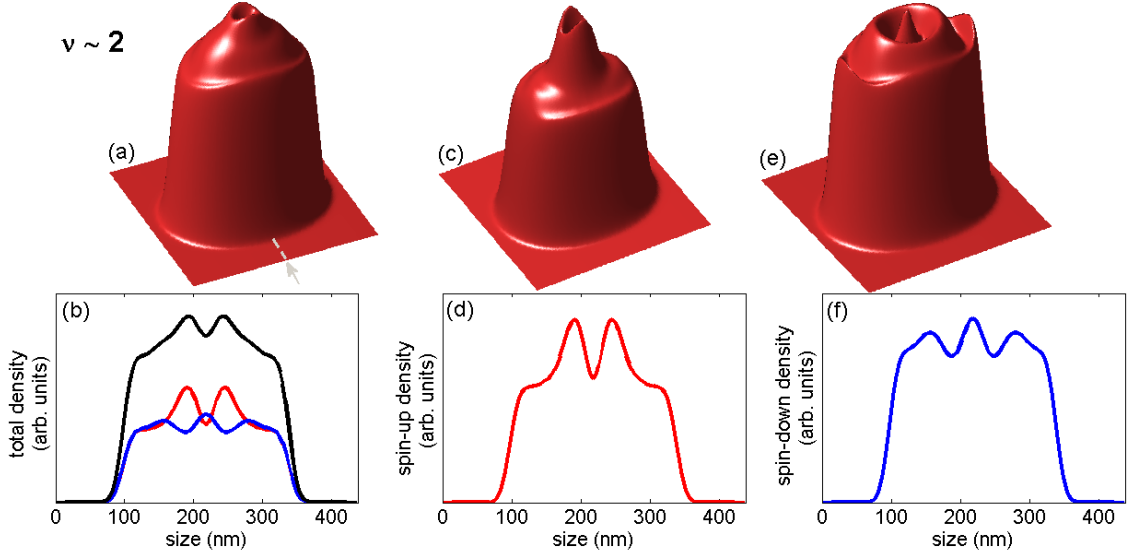}
\caption{Total and spin densities in a 48-electron QD at $B=1.492$ T
corresponding to $\nu\sim 2$. The cross sections in the lower panel
are taken along a vertical cut across the 2D densities in the upper panel.
The total spin is $S=1$.
(a-b) Total density, (c-d) spin-up density, (e-f) spin-down density.}
\label{figure4}
\end{figure}

In Figure \ref{figure4} we show the total and spin densities in
a 48-electron QD at $B=1.492$ T corresponding to $\nu\sim 2$.
The ground-state total spin is $S=1$ as a slight deviation from
an ideal $\nu=2$ state, which is fully spin-compensated, i.e.,
the states in the lowest LL are doubly filled with spin-up and 
spin-down electrons \cite{label7}, 
In our case, it seems clear in Figure \ref{figure4}(b) 
that the polarized electrons are located close to the core of the 
QD in the second-lowest Landau level (1LL). Nevertheless, the
total density is relatively flat in accordance with the $\nu=2$ state
in a parabolic QD. We note that the rotational symmetry is broken
due to the non-circular confining potential. The cross sections
in the lower panel of Figure \ref{figure4} are taken along a vertical 
cut across the 2D densities in the upper panel.

\begin{figure}
\includegraphics[width=1\columnwidth]{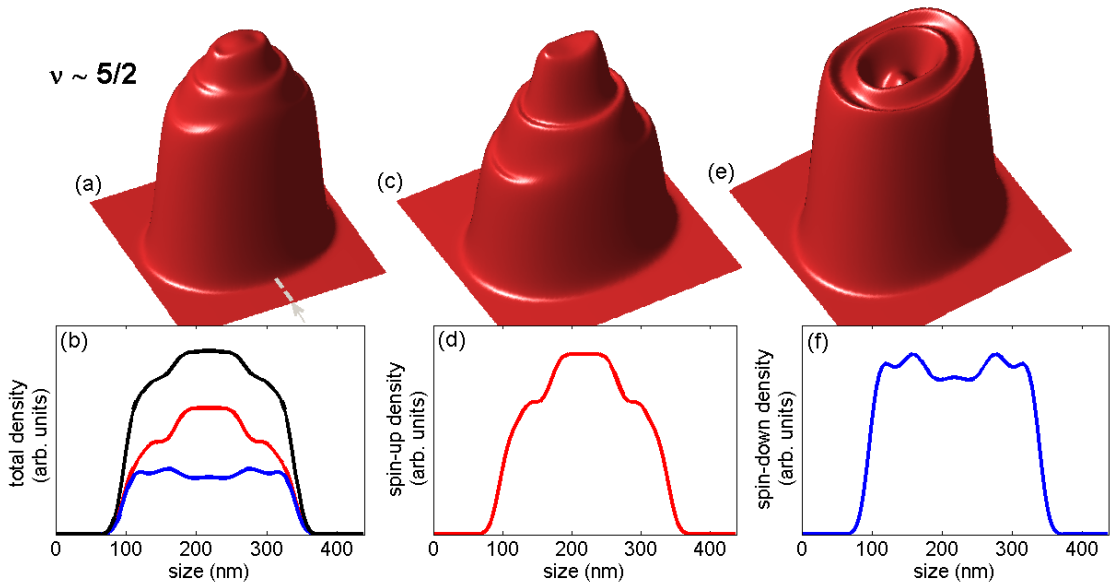}
\caption{Total and spin densities in a 48-electron QD at $B=1.15$ T
corresponding to $\nu\sim 5/2$. The cross sections in the lower panel
are taken along a vertical cut across the 2D densities in the upper panel.
The total spin is $S=4$.
(a-b) Total density, (c-d) spin-up density, (e-f) spin-down density.}
\label{figure5}
\end{figure}

Figure \ref{figure5} shows the total and spin densities similarly to
the previous case but now at $B=1.15$ T corresponding to 
$\nu\sim 5/2$. The total spin is $S=4$. It is interesting to
note that the spin polarization is strongly concentrated close to
the core of the QD. This is confirmed by sorting the Kohn-Sham states
according to their spin and angular momenta. Hence, it is clear
that the (eight) polarized electrons occupy the 1LL. We may thus call
this state as a spin droplet. Its characteristics are strikingly 
similar to those in a parabolic QD. Despite the non-circular 
potential, the density of the spin droplet in the core region 
[Figures \ref{figure5}(c) and (d)] is very smooth. 
Most likely, this is a consequence of the screening
of the irregularities in the potential by the 0LL electrons, so
that the 1LL with the spin droplet has a smooth surrounding potential.
It might be expected that the spin droplet could survive in even
more irregular geometries, i.e., in the vicinity of a quantum point
contact. This aspect of universality deserves more investigations.

An important aspect in the formation of a spin droplet is the
sufficient $N$ for the phenomenon to occur. In parabolic QDs 
spin droplets at $2\leq\nu\leq 3$ were found when
the number of electrons was $N \geq 30$. 
According to our calculations there is no considerable change in the 
critical $N$ in a non-circular geometry, so that the droplet emerges 
at $N\approx30$ as well. As discussed in Sec. \ref{introduction}
the spin polarization of the highest occupied LL is a collective
effect resulting from a high density of states close to the 
Fermi energy. This is the case if (i) $N$ is sufficiently large and
(ii) a proper fraction of the electrons occupy the highest 
LL (for $\nu\sim 5/2$ the highest LL is the second one).
The effect resembles Hund's rule: at the expense of kinetic
energy, spin polarization saves exchange energy close to a
point of degenerate states. Here, with dozens of electrons involved
in the process, the phenomenon is similar to the Stoner 
effect \cite{label23}, which predices the emergence of 
ferromagnetic alignment in a correlated electron system if
the degeneracy is high close to the Fermi level.

\subsection{Chemical potentials in the spin-droplet regime}

Finally we consider the chemical potentials to assess
the signals that the formation of spin droplets in 
leave in spin-blockade oscillations. 
The chemical potential is defined as 
$\mu(N)=E_{\rm tot}(N)-E_{\rm tot}(N-1)$. Here we consider
$\mu(48)$ by computing the energies of different spin states for $N=47$ and 
$N=48$. For both systems, respectively, the lowest energy 
state (ground-state) was chosen to calculate the chemical potential.

\begin{figure}
\includegraphics[width=1\columnwidth]{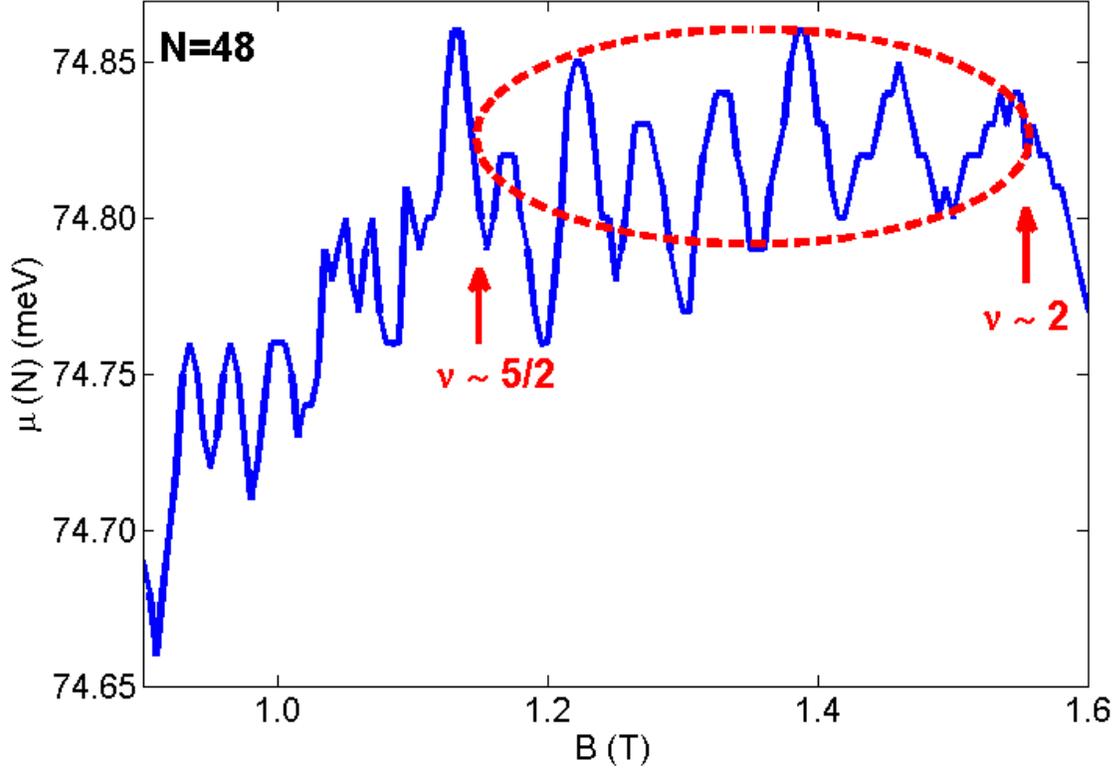}
\caption{Chemical potential for a 48-electron QD as a function
of the external magnetic field. The spin-droplet regime is
marked with a dashed ellipse.}
\label{figure6}
\end{figure}

Figure \ref{figure6} shows the chemical potential as a function
of the magnetic field in the vicinity of the spin-droplet range. 
We find a clear ``plateau'' region at $2\leq\nu\leq5/2$
superimposed by distinctive sawtooth-like oscillations. 
The structure is similar to the Coulomb-blockade peak
oscillations measured in both lateral and vertical 
QDs~\cite{label7, label8, label9}. Moreover,
the qualitative features are similar to the previous SDFT calculations 
for parabolic QDs \cite{label7}, although in this case
we find more irregularities. In fact, the irregularities in $\mu$
resemble the experimental data; it might be expected that in
actual QDs the impurities etc. induce similar effects found
here for a non-circular qH device. 

%=================================================0

\section{Summary}\label{summary}

In summary, we have studied the spin-droplet formation
in realistic quantum Hall devices in magnetic fields corresponding
to filling factors $2\leq\nu\leq 3$. We carried out self-consistent
electrostatic calculations for a GaAs/AlGaAs heterostructure with 
experimental parameters. In this way we obtained a confining potential
for electrons that can be trapped inside the quantum dot.
Our spin-density-functional-theory studies for $\sim 48$ interacting
electrons in the determined non-circular potential show that 
(i) spin polarization occurs at $\nu\sim 5/2$, (ii) polarized electrons are located at the core of the dot in the second Landau level, (iii) the spin droplet is very smooth and thus similar to that in an ideal parabolic quantum dot, (iv) the critical $N$ for the formation of the spin droplet is not affected by having a realistic (non-symmetric) potential, and (v) the spin-droplet formation leaves a signal to the chemical potential that resembles the available experimental spin-blockade data.

We hope that the present study encourages further experimental 
studies on the subject. In particular, it would be important to
detect the spin-droplet {\em directly} in an experiment
by appropriate spin-imaging methods.
In addition, although the present study confirms the stability of 
the spin droplet in a non-circular geometry, the 
state is still to be found in a large system with $N\sim 1000$
confined electrons, where it can be assumed to be extremely stable.
The ongoing experiments on quantum point contacts are likely
to bring answers to these assumptions in near future, and they may 
open up the path for further applications.

\ack
We are grateful to V. J. Goldman for providing us 
with the layer structure. This work was supported by the
Academy of Finland (HA and ER), Wihuri Foundation (ER),
ERASMUS Internship Programme (HA), the scientific 
and technological research council of Turkey (T\"UB\.ITAK) 
under grant no:TBAG-109T083 (HA and AS), IU-BAP:6970 (AS) and
Institute of Theoretical and Applied Physics (ITAP) in
Turun\c c, Turkey. CSC Scientific Computing Ltd.
is acknowledged for computational resources.

\end{document}